\documentclass[aps,prl,twocolumn,superscriptaddress]{revtex4}
\usepackage{graphicx}
\usepackage{amsmath}
\usepackage{amssymb}
\usepackage{colordvi}
\usepackage{mathrsfs}
\usepackage{bm}
\usepackage{verbatim}
\usepackage{dcolumn}
\usepackage{epsfig}
\usepackage{subfigure}
\usepackage{ulem}

\begin{document}
\title{Mesoscopic Electronic Transport in Twisted Bilayer Graphene}
\author{Yulei Han}
\affiliation{ICQD, Hefei National Laboratory for Physical Sciences at Microscale, University of Science and Technology of China, Hefei, Anhui 230026, China}
\affiliation{CAS Key Laboratory of Strongly-Coupled Quantum Matter Physics, and Department of Physics, University of Science and Technology of China, Hefei, Anhui 230026, China}
\author{Yafei Ren}
\affiliation{ICQD, Hefei National Laboratory for Physical Sciences at Microscale, University of Science and Technology of China, Hefei, Anhui 230026, China}
\affiliation{CAS Key Laboratory of Strongly-Coupled Quantum Matter Physics, and Department of Physics, University of Science and Technology of China, Hefei, Anhui 230026, China}
\author{Xinlong Dong}
\affiliation{ICQD, Hefei National Laboratory for Physical Sciences at Microscale, University of Science and Technology of China, Hefei, Anhui 230026, China}
\affiliation{CAS Key Laboratory of Strongly-Coupled Quantum Matter Physics, and Department of Physics, University of Science and Technology of China, Hefei, Anhui 230026, China}
\affiliation{School of Chemistry and Materials Science, Shanxi Normal University, Linfen, Shanxi 041004, China}
\author{Junjie Zeng}
\affiliation{ICQD, Hefei National Laboratory for Physical Sciences at Microscale, University of Science and Technology of China, Hefei, Anhui 230026, China}
\affiliation{CAS Key Laboratory of Strongly-Coupled Quantum Matter Physics, and Department of Physics, University of Science and Technology of China, Hefei, Anhui 230026, China}
\author{Wei Ren}
\affiliation{International Center for Quantum and Molecular Structures, Materials Genome Institute, Physics Department, Shanghai University, Shanghai 200444, China}
\author{Zhenhua Qiao}
\email[Correspondence author:~]{qiao@ustc.edu.cn}
\affiliation{ICQD, Hefei National Laboratory for Physical Sciences at Microscale, University of Science and Technology of China, Hefei, Anhui 230026, China}
\affiliation{CAS Key Laboratory of Strongly-Coupled Quantum Matter Physics, and Department of Physics, University of Science and Technology of China, Hefei, Anhui 230026, China}
\date{\today}
	
\begin{abstract}
  We numerically investigate the electronic transport properties between two mesoscopic graphene disks with a twist by employing the density functional theory coupled with non-equilibrium Green's function technique. By attaching two graphene leads to upper and lower graphene layers separately, we explore systematically the dependence of electronic transport on the twist angle, Fermi energy, system size, layer stacking order and twist axis. When choose different twist axes for either AA- or AB-stacked bilayer graphene, we find that the dependence of conductance on twist angle displays qualitatively distinction, i.e., the systems with ``top" axis exhibit finite conductance oscillating as a function of the twist angle, while the ones with ``hollow'' axis exhibit nearly vanishing conductance for different twist angles or Fermi energies near the charge neutrality point. These findings suggest that the choice of twist axis can effectively tune the interlayer conductance, making it a crucial factor in designing of nanodevices with the twisted van der Waals multilayers.
\end{abstract}

\maketitle

\textit{Introduction---.} The discovery of graphene~\cite{A. K. Geim1} booms extensive investigation on two-dimensional van der Waals materials, among which the bilayer graphene is considered as an ideal prototype system~\cite{A. K. Geim2}. Both experiments\cite{J. Hass1} and theoretical studies via either atomistic simulations~\cite{Latil,Shallcross1,Morell,Laissardiere,Jeil,Sboychakov,Oshiyama,Fang,Trambly1,Andelkovic,Po1,Song1,Ahn1} or low-energy continuum models~\cite{Lopes1,Mele1,Shallcross2,Bistritzer2,Lopes2,Koshino1,Koshino2,Ray1,Liu,Mele2} reveal the dramatic modification of interlayer van der Waals interactions and band structures by a twist between two graphene monolayers.
For large twist angles, the low-energy behavior of twisted bilayer graphene (tBLG) matches that of monolayer graphene, indicating the suppressed coherent interlayer electronic transport~\cite{Latil,Shallcross1,Morell,Laissardiere,Lopes1,Shallcross2,Lopes2}. For small twist angles, e.g., $\theta<10^\circ$, the suppression of Fermi velocity becomes pronounced~\cite{Shallcross1,Morell,Laissardiere,Lopes1,Shallcross2,Bistritzer2}.
When $\theta$ approaches to $1^\circ$, the strong interlayer coupling may result in more exotic phenomena, e.g., moir\'{e} flat band~\cite{Bistritzer2}, van Hove singularities~\cite{Luican1,Luican2,Lopes2}, and strongly correlated electron states~\cite{Cao1,Cao2,Sharpe_mag,Lu_mag}.
Generally, the electronic transport is a effective way to investigate the dependence of electronic properties on the twist angle.~\cite{Cao1,Cao2,Sharpe_mag,Lu_mag,Cao0,Kim0,Rickhaus,Huang0,Chung0,Bistritzer1,V. Perebeinos1,Kim}.

\begin{figure}
  \includegraphics[width=8cm]{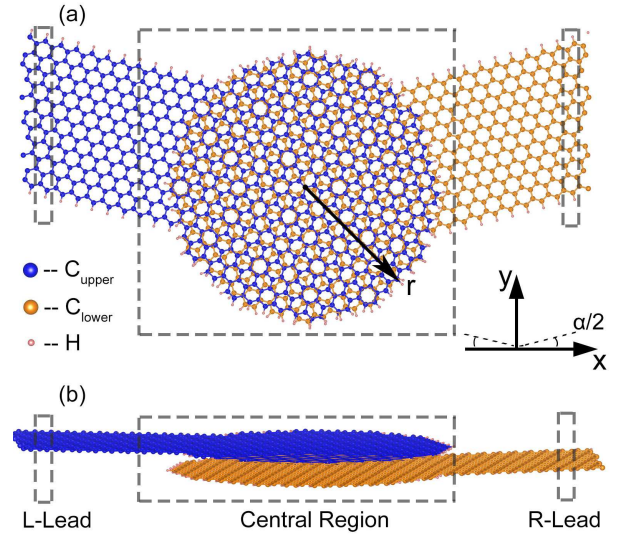}
  \caption{(a)-(b): top and side views of mesoscopic twisted bilayer graphene device in our calculation. The left (right) terminal is connected seamlessly to the upper (lower) layer. Carbon (C) atoms in blue and orange are at upper and lower layers, respectively. Hydrogen atoms in red saturate the open boundaries. ``r" is the radius of bilayer graphene disk. ``$\alpha$/2" measures the relative twist of each layer with $x$ axis and $\alpha$ thus denotes the relative twist between upper and lower layers.}\label{Fig1}
\end{figure}

Theoretically, the electronic properties of tBLG have been explored by employing atomistic calculations or low-energy continuum model~\cite{J. Hass1,Latil,Shallcross2,Lopes1,Mele1,Mele2,Bistritzer2}. The former approach is usually restricted to limited relative large commensurate angles~\cite{J. Hass1,Latil} since the atomic number of each tBLG supercell changes drastically and is easy to exceed tens of thousands for small twist angles~\cite{Shallcross2}. The later one, however, is restricted to rather small angles and the parameters used in these long wavelength theories are not unique~\cite{Lopes1,Mele1,Mele2,Bistritzer2}.
Moreover, compared to above studies on commensurate angles, the incommensurate cases are usually ignored due to the assumption of vanishing interlayer coupling. Interestingly, recent experiments reported the observation of strong interlayer coupling in an incommensurate tBLG~\cite{Yao_incomm,HeLin}.
Therefore, it is still challenging in uncover the emergent electronic properties of tBLG with arbitrary twist.
The aforementioned difficulties arise naturally in the investigation of bulk tBLG systems due to the presence of translational symmetry or large supercell. However, in mesoscopic tBLG systems, these difficulties can be avoided and the twist axes or twist angles can be manipulated easily.

In this Letter, we numerically explore the interlayer transport properties of two mesoscopic discs of graphene with a twist over a wide range of angles, which include both commensurate and incommensurate angles. By studying the dependence of the electronic transport properties on the system size, stacking order, twist axis, twist angle, and Fermi energy, we find that the twist axis plays crucial role in determining the interlayer tunneling. When the twist axis going through the ``top" site of bilayer graphene, the system is mostly conductive near the charge neutrality point and the conductance shows strong oscillate as a function of twist angle. When the twist axis going through the ``hollow" site, the system displays insulating characteristic, i.e. the conductance is nearly vanishing for Fermi energies around the charge neutrality point. By displaying the conductance as functions of Fermi energy and twist angle at different twist axes, we present a clear and complete picture on how the twist axis and the twist angle can be used to tune the interlayer conductance in mesoscopic tBLG system. This will definitely shed light on designing nanodevices from twisted van der Waals materials.

\textit{Model and Methods---.} In our calculation, a two-terminal mesoscopic device as displayed in Fig.~\ref{Fig1} is adopted, where Figs.~\ref{Fig1}(a) and \ref{Fig1}(b) display schematically the top and side views of our considered system. The central region is chosen to be two graphene layers of disk shape to avoid the variance of the central overlap area during the continuous twist. The left/right terminal is seamlessly integrated to the upper/lower layer. The dangling bonds are saturated with hydrogen (H) atoms. The terminal width is set to be 60\% of the disk diameter. Due to the computational capacity~\cite{computationalCapacity}, the disk radii are chosen to be $r=7.8,~10.7,~14.2,~20.9$ and $25.6~{\rm {\AA}}$ in our calculations. The corresponding numbers of carbon (C) and H atoms are respectively (150-C, 35-H), (270-C, 46-H), (480-C, 63-H), (1000-C, 92-H), and (1560-C, 112-H) in the central region. The interlayer distance is set to be $3.4~\rm{\AA}$. $\alpha$ indicates the relative twist between upper and lower layers.

The lattice structure of central region with connected terminals is optimized by using Vienna $ab$ $initio$ simulation package (VASP)~\cite{vasp1,vasp2,vasp3}. The two-terminal conductance is obtained by employing the NanoDcal package~\cite{nanodcal1,nanodcal2}, which combines the nonequilibrium Green's function with density functional theory. The single zeta polarized atomic orbital basis is used. In both VASP and NanoDcal calculations, the exchange correlation potential assumes the Perdew-Burke-Ernzerhof generalized gradient approximation~\cite{pbe}. Details of the conductance calculation can be found in Supplemental Materials~\cite{SM}. To provide a systematic study on the dependence of interlayer transport on the relative twist angles, we choose 61 integer angles ranging from $0^\circ$ to $60^\circ$ and 12 commensurate angles~\cite{SM}. Hereinbelow, in our considered systems, we use the two-terminal conductance to effectively reflect the interlayer tunneling.
\begin{figure}
	\includegraphics[width=8.5cm]{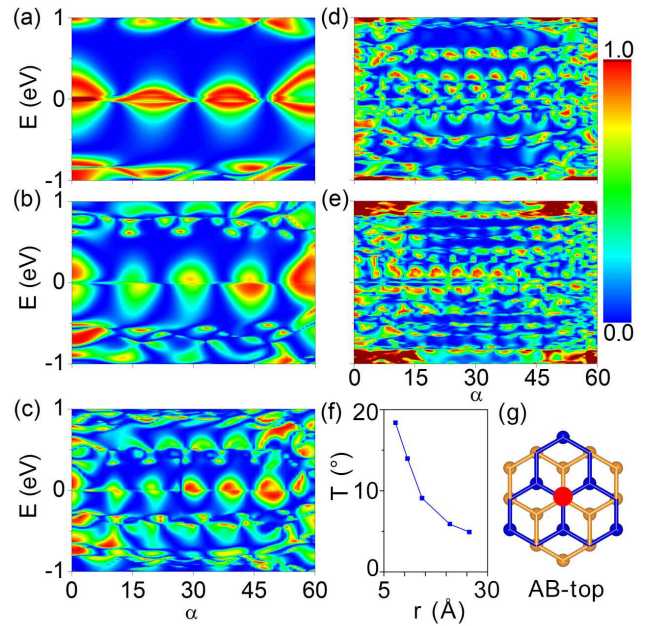}
	\caption{(a)-(e): The colormaps of two-terminal conductance $G$ in ($\alpha$, $E$) plane for twist axis shown in (g). The corresponding disk radii are $r=7.8~{\rm {\AA}}$~(a), $10.7~{\rm {\AA}}$~(b), $14.2~\rm {\AA}$(c), $20.9~{\rm {\AA}}$~(d), and $25.6~{\rm {\AA}}$~(e), respectively. Color bar is used to measure the strength of the conductance. (f): The period $T$ of conductance oscillation along with twist angle for different disk radius. (g): Schematic plot of the twist axis going through the top site (red spot) of the AB-stacked bilayer graphene.}\label{Fig2}
\end{figure}

\begin{figure*}
	\includegraphics[width=16cm]{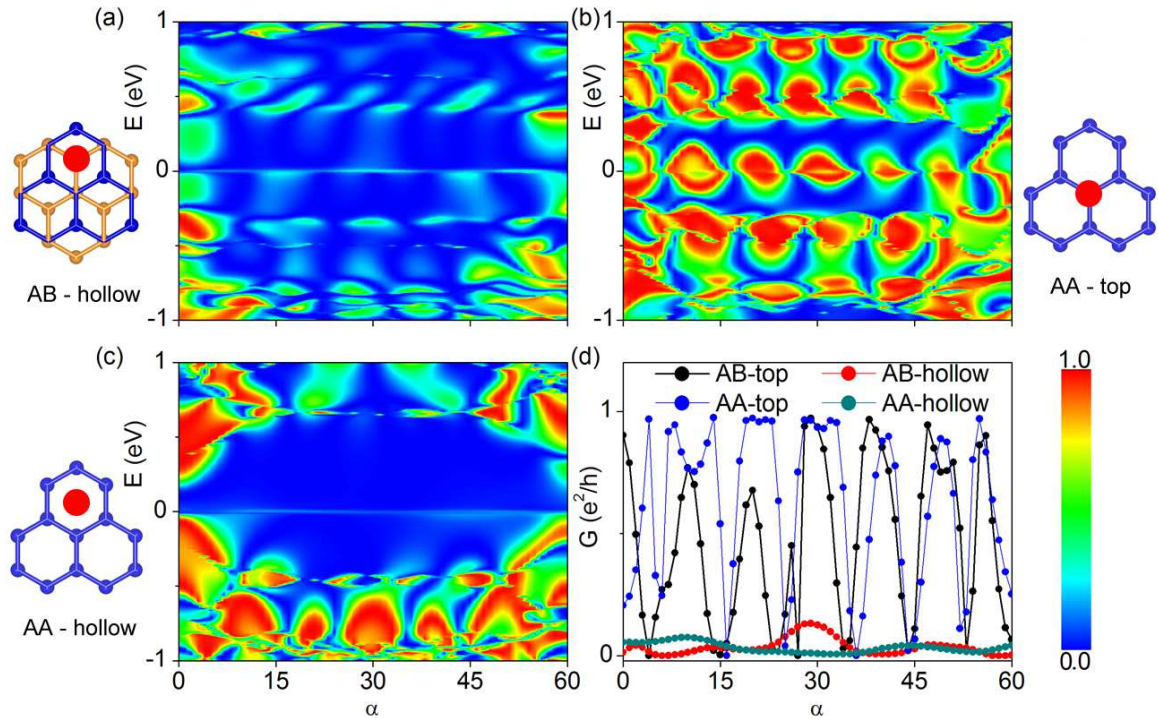}
	\caption{(a)-(c): The colormaps of the two-terminal conductance $G$ in ($\alpha$, $E$) plane for four different twist axes shown at the left side of each panel at a fixed radius of $r=14.2~{\rm {\AA}}$. Color measures the conductance. (a): ``AB-hollow". The vertical twist axis go through the hollow site (red spot) of the AB-stacked bilayer graphene. (b) and (c): ``AA-top" and ``AA-hollow". The vertical twist axes respectively go through the top (b) and hollow (c) sites of the AA-stacked bilayer graphene. (d): The conductance $G$ as a function of twist angle $\alpha$ for four different twist axes at $E=0.02~\rm {eV}$.}\label{Fig3}
\end{figure*}

\textit{Effect of System Size---.} We first study the influence of system size on the two-terminal conductance, by taking the example of the twist axis of the ``AB-top'' as schematically displayed in Fig.~\ref{Fig2}(g). The colormaps of conductance ($G$) in the space spanned by twist angle $\alpha$ and Fermi energy $E$ are displayed in Figs.~\ref{Fig2}(a)-\ref{Fig2}(e) for disk radii of $r=7.8~{\rm {\AA}}$, $10.7~{\rm {\AA}}$, $14.2~\rm {\AA}$, $20.9~{\rm {\AA}}$, and $25.6~{\rm {\AA}}$, respectively. At relative small radius [see Fig.~\ref{Fig2}(a)], the colormap of conductance is well separated into strong- and weak-conducting regions as shown in red and blue, separately. We find that conductance $G$ varies quasi-periodically as the twist angle $\alpha$ at fixed Fermi energy $E$. Specifically, at $E=0.1$~eV, the local maximum of $G$ appear at twist angles of $\alpha = 5^\circ, ~20^\circ, ~39^\circ$, and $52^\circ$, indicating relatively strong interlayer tunneling, whereas $G$ reaches its local minima that are vanishingly small at $\alpha= 13^\circ, 29^\circ$, and $47^\circ$, indicating the prohibit of tunneling from upper to lower layers. Similarly, the conductance $G$ also fluctuates from zero to non-zero as Fermi energy at fixed twist angle $\alpha$. When the Fermi energy is away from charge neutrality point gradually, the regions near maxima evolve into conducting islands surrounded by weak-conducting region shown in blue color. Around $E=0.60$ eV, the interlayer conductance becomes vanishingly small in the whole range of $\alpha$ from $0^\circ$ to $60^\circ$. These results suggest that at certain twisting angles, one can effectively tune the interlayer tunneling by electrical gating, which can be utilized to realize the ``ON" or ``OFF" functions in logic devices.

When the disk radius increases, the quasi-periodic oscillation of $G$ preserves as displayed in Figs.~\ref{Fig2}(b)-\ref{Fig2}(e) where the periodicity $T$ decreases as the system size increases. The number of regions reaching conductance maxima increases with the radius, i.e., there are 4 regions of conductance maxima near charge neutrality point at $r=7.8~\rm {\AA}$ [see Fig.~\ref{Fig2}(a)], while there are 5 regions at $r=10.7 \rm {\AA}$ [see Fig.~\ref{Fig2}(b)]. We can find that the regions of vanishingly small conductance always exist as the increase of system radius, indicating the better control of ``ON" and ``OFF" functions by tuning either the Fermi energy $E$ or the twist angle $\alpha$.
In Fig.~\ref{Fig2}(f), we fit the quasi-period $T$ of conductance oscillation along with the twist angle at different disk radii. One can see that as the radius increases, the periodicity $T$ first rapidly decreases, and then tends to converge at larger radius. Besides the integer twist angles, we also find that the transition of conductance between commensurate and incommensurate angles is smooth due to the absence of moir\'{e} periodic potential in such mesoscopic tBLG system~\cite{SM}.

\textit{Role of Twist Axis---.} Now, we move to explore the influence of the twist axis on determining the interlayer coupling. From the symmetric point of view, there are three kinds of twist axis: (i) ``AB-hollow" as displayed in Fig.~\ref{Fig3}(a); (ii) ``AB-top" or ``AA-top" in Fig.~\ref{Fig2}(f) and \ref{Fig3}(b); (iii) ``AA-hollow" in Fig.~\ref{Fig3}(c). In below, we set the system size to be $r=14.2~{\rm {\AA}}$. By comparing the conductance in the three different twist axes as shown in Figs.~\ref{Fig3}(a)-\ref{Fig3}(c), one can clearly find that, the conductances for ``AB-hollow" and ``AA-hollow" geometries are vanishingly small at Fermi energies near the charge neutrality point as twist angles deviating from zero. In a sharp contrast, the conductances for ``AB-top" and ``AA-top" for Fermi energies near the charge neutrality point are finite and oscillating in a quasi-periodic manner as a function of the twist angle. Figure~\ref{Fig3}(d) presents a direct comparison of the conductances among four different twist axes at the Fermi level of $E=0.02eV$: the systems with ``hollow" type axis exhibit nearly vanishing conductance, while the systems with ``top" type axis exhibit oscillating conductance, with the maxima approaching quantization.

The above observation of qualitative difference strongly suggests that the twist axis plays a crucial role in determining the interlayer transport, which suggests the dependence of electronic properties of tBLG on the twisting axis. This finding sheds great light in engineering ``ON/OFF" functions in nanometer sized twist-related devices. To confirm our findings based on the first principles calculation, we also construct tight-binding models for the same systems employed above to study the electronic transport properties. We find that the results from tight-binding models are qualitatively consistent with that from first principles calculation. Nevertheless, near the centers of the high-conductive regions for the ``top" axes, the tight-binding model results show vanishing conductance, which is qualitatively different from the first-principles calculations indicating some information missing in tight-binding model~\cite{SM}.

\textit{Summary---.} We studied the interlayer transport properties of two mesoscopic graphene disks with a wide range of twist angles. By employing nonequilibrium Green's function method combining with density functional theory and tight-binding models, we explore systematically the dependence of interlayer conductance on the system size, stacking order, twist axis, twist angle, and Fermi energy. By calculating the conductance with different disk radii, we find that interlayer tunneling quasi-periodically exists as a function twist angle in such mesoscopic twist bilayer graphene system. We then find that twist axes has significant influence on the electronic transport properties, i.e., the systems with ``top" twist axis displays finite conductance oscillating as a function of twist angle, while for ``hollow" twist axis, the conductance nearly vanishes for Fermi energy near the charge neutrality point. Moreover, at fixed twist angles, the conductance can be effectively controlled by tuning the Fermi energy, indicating the electrical tunable conductance in nano-devices of tBLG system. These findings imply that incommensurate angles and twist axes can greatly modify the electronic properties in mesoscopic van der Waals materials and illuminate the way to design twist multilayer nanodevices.

\textit{Acknowledgments--.} We are grateful to Profs. Qian Niu, Kai Chang, Yin Wang, and Bin Wang for useful discussions. This work was financially supported by the National Key Research and Development Program (2017YFB0405703), the National Natural Science Foundation of China (11474265 and 11974327), Anhui Initiative in Quantum Information Technologies, and the Fundamental Research Funds for the Central Universities. We are grateful to AMHPC and Supercomputing Center of USTC for providing the high performance computing resources.

%


\begin{thebibliography}{99}
\bibitem{computationalCapacity}
The maximum memory in our server is 1TB/node. In practical, the instantaneous maximum memory of a disk radii with $r=25.6~\rm{\AA}$ can be about 900 GB and it will take about 3 weeks to finish the self-consistent calculation.
\end{thebibliography}
\end{document}